\documentclass[letterpaper, journal, twoside]{IEEEtran}

\IEEEoverridecommandlockouts   

\usepackage{dsfont}
\usepackage[T1]{fontenc}
\usepackage{graphicx, stfloats} 
\usepackage{svg}
\usepackage{geometry}
\usepackage{standalone}
\usepackage{hyperref}
\usepackage{pgfplots}
\usepackage{booktabs}
\usepackage{subcaption}
\usepackage{amssymb, amsmath}
\usepackage{algorithm}
\usepackage{algpseudocode}
\usepackage{makecell}
\usepackage{bbm}
\usepackage{diagbox, multirow} 
\usepackage{color,soul} 
\usepackage{siunitx}
\usepackage{xurl}

\definecolor{vandeusen}{RGB}{73,92,111}
\definecolor{cordovan}{RGB}{152,68,71}
\definecolor{alizarin}{rgb}{0.82, 0.1, 0.26}
\definecolor{azure}{rgb}{0.0, 0.5, 1.0}

\title{Optimized Renewable Energy Planning MDP for Socially-Equitable Electricity Coverage in the US}

\author{ Riya Kinnarkar$^{1}$, Mansur M. Arief$^{2\star}$%

\thanks{$^1$Riya Kinnarkar is with the Westminster High School, GA, USA}%
\thanks{$^2$Mansur M. Arief is with the Department of Aeronautics and Astronautics, Stanford University, Stanford, CA, USA (\url{mansur.arief@stanford.edu})}%
\thanks{$^{\star}$Corresponding Author}%
\thanks{Code available at: \url{https://github.com/mansurarief/EnergyProjectMDP}}%
}

\begin{document}

\markboth{Preprint Version. August, 2025}
{Kinnarkar \& Arief (2025): Optimized Renewable Energy Planning MDP for Socially-Equitable Electricity Coverage in the US}

\maketitle

\begin{abstract}
Traditional power grid infrastructure presents significant barriers to renewable energy integration and perpetuates energy access inequities, with low-income communities experiencing disproportionately longer power outages. This study develops a Markov Decision Process (MDP) framework to optimize renewable energy allocation while explicitly addressing social equity concerns in electricity distribution. The model incorporates budget constraints, energy demand variability, and social vulnerability indicators across eight major U.S. cities to evaluate policy alternatives for equitable clean energy transitions. 
Numerical experiments compare the MDP-based approach against baseline policies including random allocation, greedy renewable expansion, and expert heuristics. Results demonstrate that equity-focused optimization can achieve 32.9\% renewable energy penetration while reducing underserved low-income populations by 55\% compared to conventional approaches. The expert policy achieved the highest reward (580$\pm$111), while the Monte Carlo Tree Search baseline provided competitive performance (527$\pm$80) with significantly lower budget utilization, demonstrating that fair distribution of clean energy resources is achievable without sacrificing overall system performance and providing ways for integrating social equity considerations with climate goals and inclusive access to clean power infrastructure.
\end{abstract}


\section{Introduction}

\IEEEPARstart{T}{he} integration of renewable energy sources such as wind and solar into existing power grids is currently hindered by significant infrastructure limitations. Traditional power grids were primarily designed to accommodate dispatchable energy sources like fossil fuels, which provide consistent and controllable output. In contrast, renewable energy sources are inherently variable and intermittent, creating a mismatch between their characteristics and the grid's capabilities \cite{sustainability_directory_2025}. This gap restricts the ability of grids to effectively manage the influx of renewable energy, presenting a major obstacle to the global transition toward cleaner energy systems.

International energy data reveal that nearly 3,000 gigawatts of renewable energy projects are awaiting grid connection, underscoring how grid capacity has become a bottleneck in the deployment of renewables \cite{iea_2023_grids}. While investments in renewable energy generation have nearly doubled since 2010, funding directed toward grid infrastructure has remained relatively unchanged, hovering around 300 billion USD annually worldwide \cite{iea_2023_grids_summary}. Experts have highlighted the urgent need for significant upgrades and modernization of power grids to support the increasing share of renewables, with targets such as achieving a net-zero power grid by 2035 in the United States \cite{nrel_2022_100_clean}. As renewable penetration grows, managing the variability through flexible resources like energy storage becomes critical; however, rendering the optimal sizing and placement of these resources is a complex optimization challenge.

Fortunately, while previous studies have shown that renewables make power grids more fragile, recent research shows the opposite, as grids with higher levels of renewable energy are actually more resilient. A comprehensive study analyzing over 2,000 blackout events across the U.S. from 2001 to 2020 found that increased renewable energy penetration correlated with fewer and shorter outages. Even during extreme weather, these grids performed better, likely because renewables are often paired with modern grid planning and interconnected systems \cite{nieman_2025_renewables_grid}. As renewables became a larger part of the energy mix, the severity of blackouts has consistently declined in terms of area affected and energy demand lost.

Fossil fuel-based power, on the other hand, is proving to be increasingly unreliable, especially under stress. For instance, during Winter Storm Elliott in 2022, the majority of power plant failures came from gas and coal, while wind energy outperformed expectations \cite{ferc_nerc_2023_elliott}. Not only can renewables meet energy demands, but they're also capable of providing essential grid services like voltage control and frequency regulation in a way that is sometimes more effective than fossil fuel plants. Studies show that solar facilities with advanced inverters can match or exceed fossil fuel plants in delivering these stability functions, reinforcing the idea that renewables are thus, not just sustainable but also dependable \cite{chang_2023_renewable_resilient}. 

Another critical aspect of the current limitations of the existing electricity grid infrastructure is its inability to effectively distribute energy generated from geographically dispersed renewable sources to areas of demand. This disparity can be seen both economically and geographically. Economically, while the average U.S. household spends about 3.5\% of its income on energy, households living in poverty often face more than twice that burden \cite{pearl_elevate_2024_energy_gap}. Unfortunately, these disparities extend to disaster response as well. An analysis of 15 million utility customers across 588 U.S. counties showed that communities with lower socioeconomic status—measured by the CDC's social vulnerability index—consistently experienced longer power outages after hurricanes. For every 10-point decrease in social standing on the index, the average outage length increased by 6.1\%, meaning residents waited roughly 170 extra minutes for electricity to return, with some areas enduring significantly longer delays \cite{ji_ganz_2024_power_outages}.

On the other hand, a promising modeling framework, named Markov Decision Process (MDP) shows promising potential in optimizing energy grids. MDP has been used in many fields, including robotics, supply chain management, and healthcare, to make sequential decisions in environments with probabilistic transitions and rewards. For example, studies have utilized an MDP framework to manage photovoltaic inverters for voltage regulation in distribution grids \cite{el2021fully}. It has also been used successfully to optimize community energy storage management with an aim to enhance social welfare in real-time electricity markets \cite{deng2020community}. In this study, we will use MDP to not only address the technical challenges of grid upgrades and energy storage placement, but also the social inequities that persist in power distribution. This is because vulnerable communities, especially those with lower socioeconomic status, consistently experience longer blackout durations after major disruptions \cite{ganz2023socioeconomic}. We will include the social disparity of blackouts, such as how low-income or vulnerable communities face longer outages, directly in the reward function, so the model doesn't just focus on cost or reliability, but also on making the grid more fair.

The remainder of this paper is structured as follows. Section~\ref{sec:related_work} reviews the current literature on renewable energy grid integration and social equity considerations in power systems, establishing the theoretical foundation for our work. Section~\ref{sec:framework} presents our MDP-based modeling framework and experimental methodology. Section~\ref{sec:experiment} examines the performance of different policy approaches across multiple U.S. cities through numerical experiments. Section~\ref{sec:discussion} discusses the policy implications of our results and their broader significance for equitable energy transitions. Finally, Section~\ref{sec:conclusion} summarizes our key findings and outlines directions for future research.

\section{Literature Review}\label{sec:related_work}

In this section, we will review the current modeling approaches for electricity grids, its resiliency, and MDP framework.

\subsection{Electricity Grid Models and Renewable Energy}

\begin{figure*}
    \centering
\includegraphics[width=0.8\linewidth]{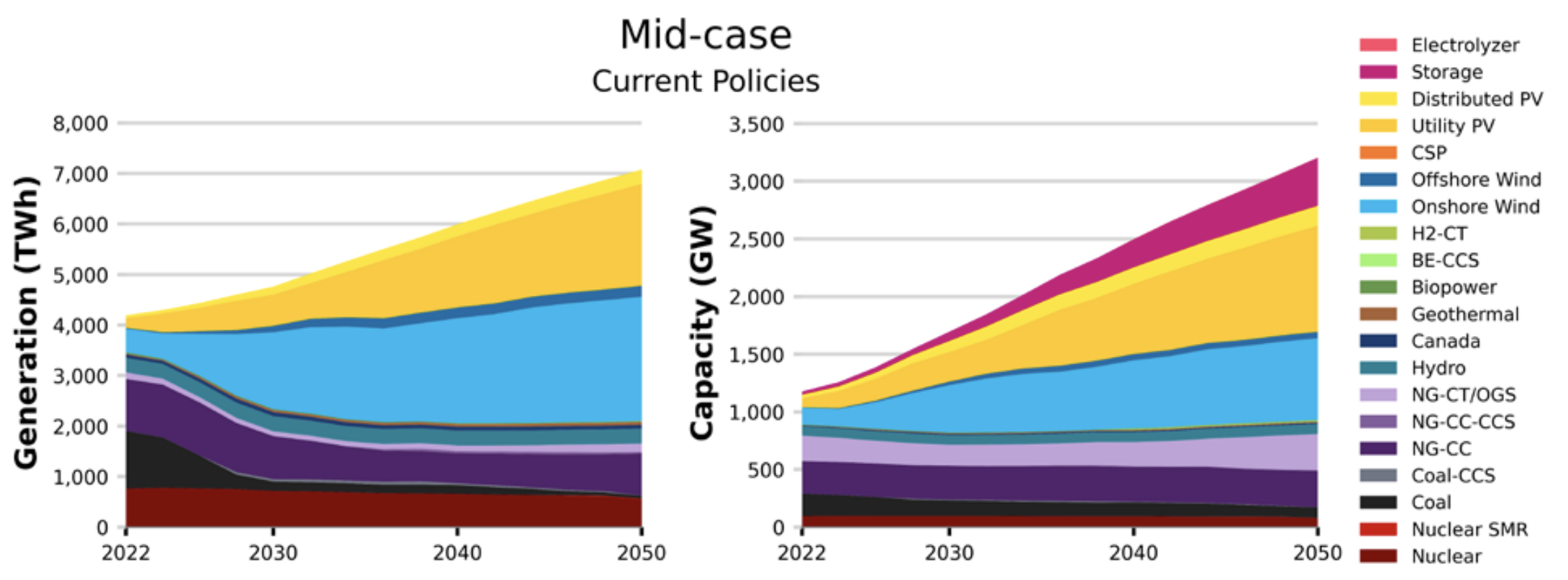}
    \caption{Current US Policies Energy Portfolio \cite{nrel_reeds}}    \label{fig:reeds_current_policies}
\end{figure*}

As renewable energy becomes a bigger part of our electricity systems, researchers have developed several ways to model and optimize grid performance. Some common approaches include capacity expansion models and mixed-integer linear programming. To begin with, capacity expansion models such as the Regional Energy Deployment System (ReEDS) look at long-term planning by simulating future grid growth based on factors like new infrastructure, demand changes, and renewable capacity \cite{nrel_reeds}. An example of current policies energy portfolio from ReEDS is illustrated in \autoref{fig:reeds_current_policies}. However, while this approach is good at showing large-scale trends and investment needs, it often assumes fixed conditions and doesn't account for day-to-day variability of the grid.

\subsection{Social Disparity of Blackouts}

While technical modeling and grid optimization have seen substantial progress, there remains a critical gap in accounting for the unequal social impacts of blackouts. Emerging research reveals that power outages disproportionately affect low-income and vulnerable communities, both in frequency and duration. A study of New York City from 2017 to 2020 found that neighborhoods with the highest social vulnerability experienced precipitation-driven outages lasting an average of 12.4 hours, which was much longer than the 7.7-hour average in the least vulnerable areas \cite{flores_casey_2024_power_outages}. Additionally, an analysis of Texas's 2021 Winter Storm Uri showed that census tracts with lower income levels and higher percentages of Hispanic residents faced extended power losses. The study noted moderate spatial inequality across both planned and weather-induced outages, reinforcing the urgent need to integrate equity into energy resilience planning \cite{coleman2023energy}.

However, in spite of this well-documented disparity, most existing modeling approaches, including those designed to strengthen smart grid integration, have not meaningfully incorporated social equity concerns. A recent review of data driven grid resilience strategies highlights this oversight, pointing out that many proposed solutions lack frameworks for evaluating their broader socio-economic impact. The authors emphasize that future research must address not only technical and regulatory challenges but also equity issues to ensure smart grid systems benefit all communities equally \cite{zhao2024optimizing}. This gap reinforces the importance of models like the one proposed in this study that account for the social disparity directly into decision-making processes.

\subsection{MDP Modeling Approaches}

MDPs have become increasingly useful to modeling decision-making under uncertainty within power systems, especially with the incorporation of renewables. Their capacity to handle stochastic dynamics over time makes them well-suited for applications where energy demand, generation, and storage are subject to variability. For instance, a study by Wang et al. introduced an energy optimization method for microgrids based on an uncertainty-aware Deep Deterministic Policy Gradient. In this study, the approach effectively addressed the challenges posed by the randomness and intermittency of renewable energy sources, which improved microgrid operations \cite{wang2025energy}.

\section{Methodology}\label{sec:framework}

In this study, we will use MDP to model equitable and sustainable energy allocation with the goal of creating optimal decision-making under constraints of budget, energy demand, social equity, and decarbonization goals. An MDP model is defined by a few components: state, action, transition function, reward function, and discount factor. We will define each of these components in the context of our energy allocation model.

\subsection{State Space}

In our model, the state represents the current configuration of the energy system across a set of $n$ cities. Each region is described by its energy demand, renewable (RE) and non-renewable (NRE) supply levels, population, and income classification. The state also includes a global budget variable which represents the amount of money remaining for future investments.

Formally, we define the state at time $t$, denoted $s_t$, as
\begin{equation}
s_t = \left( b, [d_i, r_i, n_i, p_i, I_i]_{i \in \mathcal{I}} \right),
\end{equation}
where $b$ is the remaining budget at time $t$, $d_i$ is the energy demand in region $i$, $r_i$ is the renewable energy supply in region $i$, $n_i$ is the non-renewable energy supply in region $i$, $p_i$ is the population in region $i$, $I_i$ is the income indicator (1 = high/medium income, 0 = low income), and $\mathcal{I} = \{1, 2, \ldots, n\}$ is the set of all regions.

\subsection{Action Space}

At each decision period, the planner may choose one of several possible actions, including:
\begin{itemize}
\item adding new RE facility at region $i$: $a_{r,i}$,
\item adding new NRE facility at region $i$: $a_{n,i}$,
\item removing RE facilities at region $i$: $a_{r,i}^{-}$,
\item removing NRE facilities at region $i$: $a_{n,i}^{-}$,
\end{itemize}
for each region $i \in \mathcal{I}$. Alternatively, the planner may choose to do nothing, which is denoted by $a_0$. Thus, the action space is
\begin{equation}
a_t \in \{a_{r,i}, a_{n,i}, a_{r,i}^{-}, a_{n,i}^{-} \mid i\in \mathcal I \} \cup \{a_0\}.
\end{equation}

\subsection{Transition Function}

The transition function determines how the state evolves after an action is taken. Specifically, it describes the probability of transitioning into state $s_{t+1}$ by taking action $a_t$ at state $s_t$. It is computed by
\begin{equation}
s_{t+1} = T(s_t, a_t),
\end{equation}
where $T$ is the transition function. This transition function is governed by cost structures and capacity increments. The following parameters shape these transitions: the cost of adding a new RE facility $c_{a,r}$ (NRE: $c_{a,n}$), the cost of removing an RE facility $c_{r,r}$ (NRE: $c_{r,n}$), the operating cost for generating a unit of RE in city $i$ as $c_{o,r,i}$ (NRE: $c_{o,n,i}$), and the increase of supply for each new RE facility as $\Delta s_r$ (NRE: $\Delta s_n$).

The general transition function includes operating costs for all cities and is given by:
\begin{align}
s_{t+1} &= \begin{cases}
(b - c_{a,r} - C_{\text{op}}(i, \Delta s_r, 0), \mathbf{s}'_i) & \text{if } a_t = a_{r,i} \\
(b - c_{a,n} - C_{\text{op}}(i, 0, \Delta s_n), \mathbf{s}'_i) & \text{if } a_t = a_{n,i} \\
(b - c_{r,r} - C_{\text{op}}(i, -\Delta s_r, 0), \mathbf{s}'_i) & \text{if } a_t = a_{r,i}^{-} \\
(b - c_{r,n} - C_{\text{op}}(i, 0, -\Delta s_n), \mathbf{s}'_i) & \text{if } a_t = a_{n,i}^{-} \\
(b - C_{\text{op}}(i, 0, 0), s_t) & \text{if } a_t = a_0
\end{cases}
\end{align}
where 
\begin{equation}
    \mathbf{s}'_i = [d_j, r_j + \Delta r_i \delta_{ij}, n_j + \Delta n_i \delta_{ij}, p_j, I_j]_{j \in \mathcal{I}},
\end{equation} 
and $\Delta r_i$ and $\Delta n_i$ defined in Table~\ref{tab:action_effects}, and 
\begin{align}
    C_{\text{op}}(i, \Delta r, \Delta n) &= \sum_{j \in \mathcal{I}} ( c_{o,r,j}(r_j + \Delta r \delta_{ij}) \\
    & + c_{o,n,j}(n_j + \Delta n \delta_{ij}) ),
\end{align}
where $\delta_{ij}$ is the Kronecker delta function.

\begin{table}[!t]
\centering
\caption{Supply Changes Induced by Different Actions in the Transition Function}
\label{tab:action_effects}
\begin{tabular}{@{}llcc@{}}
\toprule
\textbf{Action Type} & \textbf{Description} & $\Delta r_i$ & $\Delta n_i$ \\
\midrule
$a_{r,i}$ & Add RE facility in $i$ & $\Delta s_r$ & $0$ \\
$a_{n,i}$ & Add NRE facility in $i$ & $0$ & $\Delta s_n$ \\
$a_{r,i}^{-}$ & Remove RE facility in $i$ & $-\Delta s_r$ & $0$ \\
$a_{n,i}^{-}$ & Remove NRE facility in $i$ & $0$ & $-\Delta s_n$ \\
$a_0$ & Do nothing (no infrastructure change) & $0$ & $0$ \\
\bottomrule
\end{tabular}
\end{table}

To reflect demand uncertainty, we assume $d_i \sim \mathcal{N}(\mu_i, \sigma_i^2)$, where $\mathcal{N}(\mu, \sigma^2)$ is the normal distribution with mean $\mu$ and variance $\sigma^2$. In this study, we assume $\mu_i$ is the baseline demand in region $i$ and $\sigma_i = 1$.

\subsection{Reward Function}

The reward function captures the multiple, and sometimes competing, objectives of the planner. In this study, we consider the following objectives:
\begin{enumerate}
\item Remaining budget (positive weight), which encourages cost efficiency. 
\item Low-income population without sufficient energy (negative weight), which penalizes inequitable access.
\item Population fully supplied by RE (positive weight), which incentivizes clean energy transitions.
\end{enumerate}

The reward function is formulated as:
\begin{align}
R(s_t, a_t) &= w_1 \cdot b_{t+1} + w_2 \cdot \sum_{i: I_i = 0} \max(0, d_i - (r_i + n_i)) \cdot p_i \nonumber \\
&\quad + w_3 \cdot \sum_{i \in \mathcal{I}} \min(r_i, d_i) \cdot p_i,
\end{align}
where $w_1 = 0.15$, $w_2 = -25$, and $w_3 = 12$ based on the experimental parameters in Table~\ref{tab:experiment_params}.

\subsection{Discount Factor}
We adopt a discount factor $\gamma = 0.95$ to reflect the time value of money and long-term planning goals.

\section{Numerical Experiments}\label{sec:experiment}

To evaluate the effectiveness of our MDP model in optimizing energy distribution under real-world constraints, we design a set of numerical experiments using synthetic but realistic data. The complete implementation, including all algorithms, policies, and evaluation frameworks, is available as open-source software at \url{https://github.com/mansurarief/EnergyProjectMDP}. We consider various cost components for electricity provision for both renewable and non-renewable energy sources, including budget, installation of new facilities, and operating costs. We also incorporate supply capacity for each new facility installed. To account for costing and economics, we consider a discount factor closer to 1 (0.95). To balance the different objectives we laid out in Section~\ref{sec:framework}, we use preset objective weights for the experiment. All these values are summarized in Table~\ref{tab:experiment_params}.

The simulation includes eight major U.S. cities (Atlanta, New York City, Houston, Phoenix, Denver, Memphis, Seattle, and San Antonio) selected to reflect diverse demographics, income levels, and energy infrastructures. The simulated characteristics of the cities (not real values) that we consider are summarized in Table~\ref{tab:city_params}.

\begin{table}[!t]
\centering
\caption{Experiment Parameters}
\label{tab:experiment_params}
\begin{tabular}{@{}ll@{}}
\toprule
\textbf{Parameter} & \textbf{Description} \\
\midrule
Initial Budget & \$3,000 million \\
Cost of Adding RE ($c_{a,r}$) & \$180 million \\
Cost of Adding NRE ($c_{a,n}$) & \$120 million \\
Cost of Removing RE ($c_{r,r}$) & \$120 million \\
Cost of Removing NRE ($c_{r,n}$) & \$180 million \\
RE Operating Costs ($c_{o,r,i}$) & \$8–14/MWh\\
NRE Operating Costs ($c_{o,n,i}$) & \$35–48/MWh\\
Supply Capacity Increment ($\Delta s_r, \Delta s_n$) & 100 GW each \\
Discount Factor ($\gamma$) & 0.95 \\
Budget Weight ($w_1$) & 0.15\\
Low-income penalty weight ($w_2$) & $-25$ \\ 
RE access weight ($w_3$) & $12$\\
\bottomrule
\end{tabular}
\end{table}

\begin{table*}[!t]
\centering
\caption{City Parameters}
\label{tab:city_params}
\begin{tabular}{@{}lcccccc@{}}
\toprule
\textbf{City} & \textbf{Population} & \textbf{\% Low-Income} & \textbf{RE Supply} & \textbf{NRE Supply} & \textbf{Energy Demand} & \textbf{Low-Income?} \\
 & ($p_i$) & & ($r_i$, GW) & ($n_i$, GW) & ($d_i$, MWh) & ($I_i = 0$) \\
\midrule
Atlanta & 498,000 & 33\% & 14 & 19 & 580 & Yes \\
New York City & 8,468,000 & 27\% & 26 & 24 & 2100 & Yes \\
Houston & 2,305,000 & 20\% & 13 & 17 & 1030 & No \\
Phoenix & 1,608,000 & 26\% & 10 & 18 & 860 & Yes \\
Denver & 727,000 & 15\% & 11 & 10 & 530 & No \\
Memphis & 628,000 & 35\% & 8 & 15 & 420 & Yes \\
Seattle & 753,000 & 12\% & 12 & 8 & 550 & No \\
San Antonio & 1,452,000 & 28\% & 9 & 16 & 900 & Yes \\
\bottomrule
\multicolumn{7}{@{}l@{}}{\footnotesize Low income classification is derived from a threshold of 25\% or more of the low-income population.}
\end{tabular}
\end{table*}

\begin{figure*}[!t]
\centering
\includegraphics[width=\textwidth]{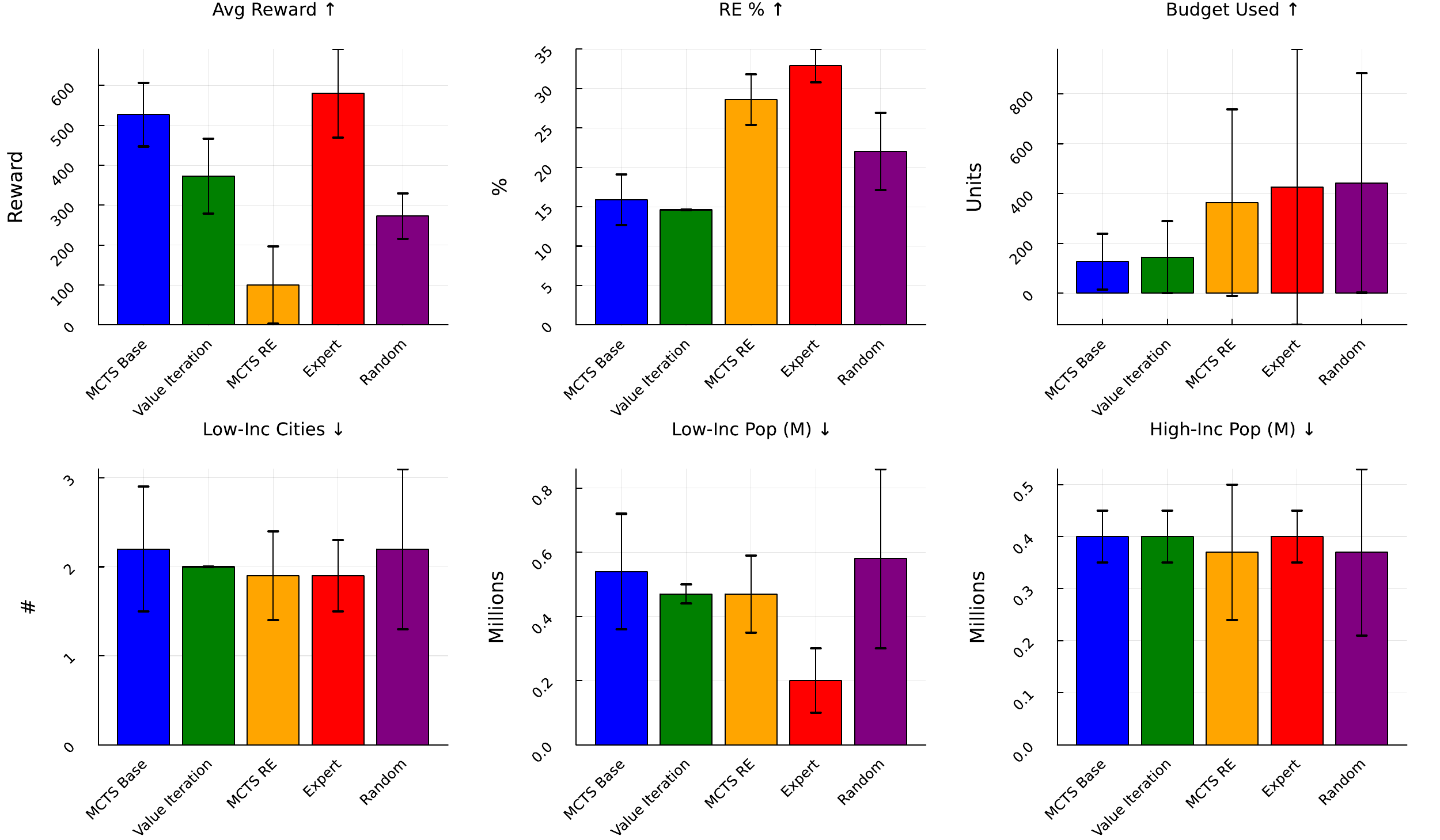}
\caption{Performance Metrics Comparison}\label{fig:performance_metrics}
\end{figure*}

We compare the MDP optimal policy with several benchmark policies: random policy (Random), expert heuristic (Expert), and MDP-optimized policy. We compare two different MDP solvers: Value Iteration and Monte Carlo Tree Search (MCTS). For MCTS, we solve for two different sets of weights, one assigning equal importance (adjusted by the objective units) which we call MCTS Base and one in which we emphasize renewable energy deployment (MCTS RE). The value iteration solver uses a discretized version of the state space, which is a common practice in MDPs. We compare these policies based on final weighted objectives such as remaining budget, equity in energy access, and renewable energy penetration. The results are summarized in Table~\ref{tab:performance}. We also plot the average discounted reward for each policy and various metrics in \autoref{fig:performance_metrics}. 


\begin{table*}[!t]
\centering
\caption{Performance Comparison}
\label{tab:performance}
\begin{tabular}{@{}lccccccc@{}}
\toprule
\textbf{Policy} & \textbf{Avg Reward} & \textbf{RE \%} & \textbf{Budget Used} & \textbf{Low-Inc} & \textbf{High-Inc} & \textbf{Low-Inc Pop} & \textbf{High-Inc Pop} \\
& \textbf{$\uparrow$} & \textbf{$\uparrow$} & \textbf{$\uparrow$} & \textbf{Cities $\downarrow$} & \textbf{Cities $\downarrow$} & \textbf{(M) $\downarrow$} & \textbf{(M) $\downarrow$} \\
\midrule
MCTS Base & 527 $\pm$ 80 & 15.9 $\pm$ 3.2 & 127 $\pm$ 112 & 2.2 $\pm$ 0.7 & 1.8 $\pm$ 0.4 & 0.54 $\pm$ 0.18 & 0.40 $\pm$ 0.05 \\
Value Iteration & 373 $\pm$ 94 & 14.6 $\pm$ 0.0 & 145 $\pm$ 14 & 2.0 $\pm$ 0.0 & 1.8 $\pm$ 0.4 & 0.47 $\pm$ 0.03 & 0.40 $\pm$ 0.05 \\
MCTS RE & 100 $\pm$ 97 & 28.6 $\pm$ 3.2 & 364 $\pm$ 374 & 1.9 $\pm$ 0.5 & 1.7 $\pm$ 0.6 & 0.47 $\pm$ 0.12 & 0.37 $\pm$ 0.13 \\
Expert & 580 $\pm$ 111 & 32.9 $\pm$ 2.1 & 426 $\pm$ 553 & 0.9 $\pm$ 0.4 & 1.8 $\pm$ 0.4 & 0.20 $\pm$ 0.10 & 0.40 $\pm$ 0.05 \\
Random & 27 $\pm$ 57 & 22.0 $\pm$ 4.9 & 442 $\pm$ 440 & 2.2 $\pm$ 0.9 & 1.6 $\pm$ 0.7 & 0.58 $\pm$ 0.28 & 0.37 $\pm$ 0.16 \\
\bottomrule
\multicolumn{8}{@{}l@{}}{\footnotesize $\uparrow$ = higher is better, $\downarrow$ = lower is better}
\end{tabular}
\end{table*}

\section{Discussion}\label{sec:discussion}

The numerical experiments in Table~\ref{tab:performance} reveal substantive differences in policy performance across multiple dimensions, offering insights into the trade-offs between efficiency, equity, and renewable energy deployment.

\subsection{Expert Policy Performance} The Expert policy achieved the highest average reward (580 $\pm$ 111), demonstrating the value of domain-informed heuristics in complex multi-objective optimization problems. This policy achieved the highest renewable energy penetration (32.9\% $\pm$ 2.1\%) while simultaneously addressing equity concerns most effectively, reducing underserved low-income populations to 0.20 $\pm$ 0.10 million—a 65\% improvement over the Random baseline (0.58 $\pm$ 0.28 million). However, this superior performance comes at significant budgetary cost (426 $\pm$ 553 million), suggesting diminishing returns at higher performance levels.

\subsection{MCTS Algorithm Analysis} The MCTS Base policy demonstrates remarkable efficiency, achieving 91\% of the Expert policy's reward while utilizing only 30\% of the budget (127 $\pm$ 112 million vs. 426 $\pm$ 553 million). This efficiency stems from MCTS's ability to explore the action space systematically while balancing immediate and long-term rewards. The algorithm's adaptability is particularly valuable—by adjusting objective weights, stakeholders can fine-tune the policy to emphasize different priorities without fundamental algorithmic changes.

The MCTS RE variant illustrates the consequences of single-objective optimization in multi-objective contexts. While achieving substantial renewable penetration (28.6\% $\pm$ 3.2\%), second only to the Expert policy, it suffers from poor budget efficiency and the lowest overall reward (100 $\pm$ 97). This demonstrates that maximizing renewable deployment without considering economic constraints leads to suboptimal system-wide outcomes.

\subsection{Value Iteration Limitations} Value Iteration's underperformance (373 $\pm$ 94 average reward) relative to MCTS methods highlights the computational challenges of discrete state space representations in high-dimensional MDPs. The low variance in its renewable energy percentage (14.6\% $\pm$ 0.0\%) suggests convergence to a local optimum, limiting its adaptability to different scenarios or preference structures.

\subsection{Equity-Efficiency Trade-offs} A critical finding is that equity improvements need not compromise overall system performance. The Expert policy achieved both the highest renewable penetration and the best equity outcomes, while MCTS Base maintained competitive equity performance (0.54 $\pm$ 0.18 million underserved population vs. Expert's 0.20 $\pm$ 0.10 million) with superior budget efficiency. This challenges conventional assumptions about inherent trade-offs between social equity and economic efficiency in infrastructure planning.

\subsection{Policy Implementation Implications} The results suggest a tiered implementation strategy: MCTS Base for resource-constrained environments where budget efficiency is paramount, and Expert-guided approaches when superior outcomes justify higher investment. The substantial performance gap between algorithmic approaches and the Random baseline (580 vs. 27 average reward) underscores the critical importance of systematic planning in renewable energy allocation.

\section{Conclusion}\label{sec:conclusion}

While traditional power grid infrastructure perpetuates energy access inequities—with low-income communities experiencing disproportionately longer outages and higher energy burdens—existing renewable energy planning models have largely failed to integrate social equity considerations into their optimization frameworks. This research establishes that Markov Decision Process frameworks can effectively optimize renewable energy allocation while simultaneously addressing social equity concerns in electricity distribution, demonstrating that fair energy access and economic efficiency are not mutually exclusive objectives. Through comprehensive testing across eight major U.S. cities using synthetic but realistic data, the study reveals three critical insights: domain-informed expert heuristics achieve the highest performance metrics (32.9\% renewable penetration and 65\% reduction in underserved populations) but require substantial budget investment, Monte Carlo Tree Search-based approaches provide exceptional efficiency-performance trade-offs (achieving 91\% of expert performance with only 30\% of the budget), and most significantly, equity improvements can be realized without compromising overall system performance—challenging conventional assumptions about inherent trade-offs between social justice and economic optimization in infrastructure planning. The results underscore the transformative potential of MDP-based decision tools for guiding sustainable and socially responsible energy transitions, while highlighting the need for future research to incorporate real-world data, additional grid constraints, and hybrid optimization strategies that combine algorithmic rigor with domain expertise to further advance both resilience and fairness in clean energy deployment.

\bibliographystyle{IEEEtran}
\bibliography{refs}

\end{document}